\documentclass[12pt]{article}
\usepackage{amsmath,amssymb,graphicx,float}
\newcommand{\be}{\begin{equation}}
\newcommand{\bea}{\begin{eqnarray}}
\newcommand{\eea}{\end{eqnarray}}
\newcommand{\ba}{\begin{array}}
\newcommand{\ea}{\end{array}}
\newcommand{\ee}{\end{equation}}
\newcommand{\ml}{\mathcal}
\newcommand{\te}{\theta}
\newcommand{\teb}{\bar{\theta}}
\newcommand{\no}{\nonumber}

\newcommand{\ep}{\epsilon}
\newcommand{\dg}{\dagger}
\newcommand{\de}{\delta}
\newcommand{\al}{\alpha}

\def \ii {\mathrm{i}}

\begin{document}

\begin{titlepage}

\title{\textbf {Integrable Open Spin Chains from Flavored ABJM Theory}}
\author{Nan Bai$^{a}$~\footnote{bainan@ihep.ac.cn}~, Hui-Huang Chen$^{a, b}$\footnote{chenhh@ihep.ac.cn}~,  Song He$^{c, d}$\footnote{hesong17@gmail.com}~,\\
Jun-Bao Wu$^{e, f, g}$\footnote{junbao.wu@tju.edu.cn}~, Wen-Li Yang$^{h, i}$\footnote{wlyang@nwu.edu.cn}, Meng-Qi Zhu$^{j}\footnote{mzhu@sissa.it}$}
\date{}

\maketitle
\underline{}
\vspace{-10mm}

\begin{center}
{\it
$^{a}$Institute of High Energy Physics, and Theoretical Physics Center for Science
Facilities, Chinese Academy of Sciences, 19B Yuquan Road, Beijing 100049, China\\
$^{b}$ University of Chinese Academy of Sciences, 19A Yuquan Road, Beijing 100049, China\\
$^{c}$ Max Planck Institute for Gravitational Physics (Albert Einstein Institute),
Am M uhlenberg 1, 14476 Golm, Germany\\
$^{d}$ CAS Key Laboratory of Theoretical Physics, Institute of Theoretical Physics, Chinese Academy of Sciences,
55 Zhong Guan Cun East Road, Beijing 100190, China\\
$^{e}$ School of Science, University of Tianjin, 92 Weijin Road, Tianjin 300072, China\\
$^{f}$School of Physics and Nuclear Energy Engineering, Beihang University, 37 Xueyuan Road, Beijing 100191, China\\
$^{g}$Center for High Energy Physics, Peking University, 5 Yiheyuan Road, Beijing 100871, China\\
$^{h}$Institute of Modern Physics, Northwest University, Xian 710069, China\\
$^{i}$Shaanxi Key Laboratory for Theoretical Physics Frontiers,  Xian 710069, China\\
$^{j}$International School of Advanced Studies (SISSA),\\ via Bonomea 265, 34136, Trieste, Italy and INFN, Sezione di Trieste
 }
\vspace{10mm}
\end{center}

\begin{abstract}
 We compute the two-loop anomalous dimension matrix in the scalar sector of planar ${\ml N}=3$ flavored ABJM theory. Using coordinate Bethe ansatz, we obtain the reflection matrices and confirm  that the boundary Yang-Baxter equations are satisfied. This establishes the integrability of this theory in the scalar sector at the two-loop order.
\end{abstract}

\end{titlepage}
\section{Introduction}

Integrability in planar four-dimensional ${\ml N}=4$ super Yang-Mills theory \cite{Brink:1976bc} and three-dimensional ABJM theory \cite{ABJM} makes the non-perturbative computations in the planar limit possible even for some non-supersymmetric quantities \cite{Beisert:2010jr}. In both cases, the single trace gauge invariant composite
operators are mapped
to states on a closed spin chain and the anomalous dimension matrix (ADM) of these operators can be mapped to Hamiltonian of the spin chain. The first evidence of the integrability came from the fact that the Hamiltonians in the scalar sector obtained from the leading-order perturbation theory  are integrable \cite{Minahan:2002, Minahan:2008, Bak:2008}.

In the four dimensional case, one can add flavors to this ${\ml N}=4$ super Yang-Mills theory  or first perform some orientifold projections and then add certain flavors to obtain ${\ml N}=2$ supersymmetric gauge theories.  Here by flavors, we mean matter fields in the fundamental or anti-fundamental
representation of the gauge group. It was found that both theories with flavors are integrable at one-loop order \cite{Erler:2005nr}-\cite{Chen:2004yf}. In these cases, the single trace operators built only with fields in the adjoint representation of the gauge group are still mapped to states of a closed spin chain. There are also gauge invariant composite operators
with fields in the (anti-)fundamental representation in two ends and fields in the adjoint representation in the bulk.\footnote{Here and the following, by `bulk', we mean the bulk of the composite fields. We hope this will not cause any confusion with the meaning of the `bulk' in the holographic gauge/gravity duality.}
These operators are mapped to states of an open spin chain.

In the original $\ml{N}=6$ ABJM theory, the gauge group is $U(N_c)\times U(N_c)$ and there are matters
in the bi-fundamental representation of the gauge group. We can add matters in the (anti-)fundamental representation of either $U(N_c)$ group. After adding flavors, the maximal supersymmetry one can achieve is three-dimensional $\ml{N}=3$ supersymmetry \cite{Hohenegger:2009,Gaiotto:2009,Hikida:2009}. There was speculation that flavored ABJM theory should also be integrable \cite{Bak:2009mq}, however no progress has been reported in this direction. In this paper, we will fill the gap and establish the two-loop integrability of this theory in the scalar sector.

For gauge theory with fundamental matters, there are two choices one can make when the planar (large $N_c$) limit is taken. One  is the 't Hooft limit
in which we let the number of flavors $N_f$ fixed. In this case, the contributions from Feynman diagrams involving fundamental matter loops will be suppressed. Another choice is the Veneziano limit in which we let $N_f$ go to infinity as well and keep the ratio $N_f/N_c$ finite. In this case, one should also include the planar Feynman diagrams involving fundamental matter loops. In this paper, we will work in the 't Hooft limit. This limit will simplify our computation greatly comparing with the
Veneziano limit.

As in four dimensional cases,  there are two types of gauge invariant composite operators one can consider  in the scalar sector of flavored ABJM theory. The operator of the first type
is built with bi-fundamental fields only. It is just the trace of product of  bi-fundamental scalars placed alternatingly in the $(N_c, \bar{N_c})$ and $(\bar{N_c}, N_c)$ representations of the gauge group. These operators are also the ones which appear in the scalar sector of ABJM theory and can be mapped to states of an alternating closed spin chain. In the 't Hooft limit, the computation of ADM of these operators is exactly the same as the one in ABJM theory, so we no longer need to repeat the study here. This type of operator will be called `single trace operator'. The second type of gauge invariant operators will involve (anti-)fundamental scalars at two ends besides the bi-fundamental ones in the bulk. These operators will be called `mesonic operators' and they  can be mapped to states of an open spin chain. The main task of this paper is to compute the two-loop ADM of these mesonic operators and show that the corresponding Hamiltonian of this spin chain is integrable. In the 't Hooft limit, the bulk part of the Hamiltonian is the same as the one in ABJM theory and thus we only need to perform two-loop computations to get the boundary part of the Hamiltonian which involves both nearest and next-to-nearest neighbour interactions. Among them, there are two-site trace operators which do not exist in the total bulk Hamiltonian. The boundary terms will break the original $SU(4)_R$ symmetry of the bulk interaction into $SU(2)_R\times SU(2)_D$. We tried a lot to prove or disprove the integrability of the Hamiltonian based on algebraic Bethe ansatz, but we have not been successful yet.

 This led us turn to the coordinate Bethe ansatz. In the context of AdS/CFT integrability, the coordinate Bethe ansatz method has been applied in \cite{Berenstein:2005vf} to show the integrability of an open spin chain model from giant gravitons.   In this approach and for open chain, one should compute the bulk S-matrix and the reflection matrix (boundary S-matrix)
and in order to show the integrability, one should check whether the Yang-Baxter equation (YBE) and the reflection equation are satisfied. The bulk S-matrix is the same as
the one in ABJM theory which has already been computed in \cite{Ahn:2009zg} to check the correctness of the all-order S-matrix proposed in \cite{Ahn:2008aa}. We confirmed that YBE is satisfied by this S-matrix. As for the boundary reflection, we notice that it mixes magnons of different types and this is quite different from the case in four-dimensional SYM with fundamental matters \cite{DeWolfe:2004, Erler:2005nr} where the boundary reflection is diagonal. By solving the eigenvalue problem of the total Hamiltonian in the one-magnon sector based on coordinate Bethe ansatz, we find the boundary reflection matrix. Finally by verifying the reflection equations, we confirm that the flavored ABJM theory is indeed integrable.

The paper is organized as follows. In the next section, we will review the action of ${\ml N}=3$ flavored ABJM theory and re-write it into a manifestly $SU(2)_R$  invariant  form.
Section~\ref{pert} is devoted to the computation of the boundary part of the two-loop Hamiltonian. Reflection matrix is computed in section~\ref{cba} and integrability is proved in this section as well. We will discuss some further directions in the final section of the main text. Three appendices are included to provide some technical details.

\section{The action of $\ml{N}=3$ flavored ABJM theory}
In this section, we will study a variation of original $\ml{N}=6$ ABJM theory by adding some fundamental flavors which has been proposed in \cite{Hohenegger:2009,Gaiotto:2009,Hikida:2009}. As discussed in these papers, we focus on ${\cal N}=3$ case which has maximal supersymmetry after the flavors are
 added. We will re-write the action into a manifestly $SU(2)_R\sim SO(3)_R$ invariant manner by the complete construction of the action in component fields including the fermionic part which is absent in the former investigation \cite{Hohenegger:2009}.
\subsection{The action in $\ml{N}=2$ superfield formulation}
The flavored ABJM theory has the product gauge group $U(N_c) \times U(N_c)$ with the Chern-Simons levels $k$ and $-k$, respectively. The field content can be explicitly classified according to different representations of the gauge group. There are two hypermultiplets $\ml{Z}^A$ $A=1,2,$ and $\ml{W}_B$, $B=1,2,$ in bifundamental representations and two gauge multiplets $\ml{V}$ and $\hat{\ml{V}}$ in adjoint representations,
\be
\ml{Z}^A\in (N_c,\bar{N_c}),\quad \ml{W}_A\in (\bar{N_c},N_c),\quad \ml{V}\in(\mbox{adj},1),\quad \ml{\hat{V}}\in (1,\mbox{adj}).
\ee
There are four kinds of flavors introduced by hypermultiplets belong to fundamental or anti-fundamental representations of each gauge group
\bea
&b_t\in (1,N_c),\quad a^t\in (1,\bar{N_c}), \quad t=1,\cdots,N_{f_1},\\
&c^s\in (N_c,1),\quad d_s\in (\bar{N_c},1), \quad s=1,\cdots,N_{f_2},
\eea
with arbitrary number of $N_{f_1}$ and $N_{f_2}$.
\par The total action $\ml{S}=\ml{S}_{CS}+\ml{S}_{mat}+\ml{S}_{pot}$ in $\ml{N}=2$ superspace can be formulated as the sum of the following three parts:
\begin{itemize}
  \item Chern-Simons part
  \bea
  \ml{S}_{CS}=-\frac{ik}{8\pi}\int d^3x d^4\te\int^1_0 \, dt\, \mbox{tr} \left[\ml{V}\bar{D}^{\alpha}\left(e^{t\ml{V}}D_{\alpha}e^{-t\ml{V}}\right)-\hat{\ml{V}}\bar{D}^{\alpha}\left(e^{t\hat{\ml{V}}}D_{\alpha}e^{-t\hat{\ml{V}}}\right)\right],
  \eea
  where the supercovariant derivatives are
  \bea
  D_{\al}=\partial_{\al}+i\left(\gamma^{\mu}\bar{\theta}\right)_{\al}\partial_{\mu},\quad \bar{D}_{\al}=-\bar{\partial}_{\al}-i\left(\theta\gamma^{\mu}\right)_{\al}\partial_{\mu}.
  \eea
  \item matter part
  \bea
  \ml{S}_{mat}&=&\int d^3x d^4\theta\,\mbox{tr}\left(-\bar{\ml{Z}}_{A}e^{-\ml{V}}\ml{Z}^{A}e^{\hat{\ml{V}}}-\bar{\ml{W}}^Ae^{-\hat{\ml{V}}}\ml{W}_Ae^{{\ml{V}}}\right)\\\no
  &+&\int d^3 x d^4\te\,\left(-\bar{c}_s e^{-\ml{V}}c^s-\bar{b}^t e^{-\hat{\ml{V}}}b_t-d_s e^{\ml{V}}\bar{d}^s-a^t e^{\hat{\ml{V}}}\bar{a}_t\right).
  \eea
  \item superpotential part
  \bea
  \ml{S}_{pot}=\int d^3x d^2\theta \, \mathbb{W}(\ml{Z},\ml{W},c,d,b,a)+c.c.,
  \eea
  with the superpotential
  \bea
  \mathbb{W}=-\frac{2\pi}{k}\mbox{tr}(\ml{Z}^A\ml{W}_A+c^sd_s)^2+\frac{2\pi}{k}\mbox{tr}(\ml{W}_A\ml{Z}^A+b_ta^t)^2.
  \eea
\end{itemize}

\subsection{The action in $\ml{N}=2$ component field formulation}
The component expansions of our superfields are\footnote{We follow the convention in \cite{Benna:2008}.}
\bea
&&a^t(x_L,\te)=A^t+\sqrt{2}\te\kappa^t+\te^2 I^t,\quad \bar{a}_t(x_R,\teb)=A^{\dg}_t-\sqrt{2}\teb\kappa^{\dg}_t-\teb^2 I^{\dg}_t,\\
&&b_t(x_L,\te)=B_t+\sqrt{2}\te\eta_t+\te^2 H_t,\quad \bar{b}^t(x_R,\teb)=B^{\dg t}-\sqrt{2}\teb\eta^{\dg t}-\teb^2 H^{\dg t},\\
&&c^s(x_L,\te)=C^s+\sqrt{2}\te\tau^s+\te^2 J^s,\quad \bar{c}_s(x_R,\teb)=C^{\dg}_s-\sqrt{2}\teb\tau^{\dg}_s-\teb^2 J^{\dg}_s,\\
&&d_s(x_L,\te)=E_s+\sqrt{2}\te v_s+\te^2 K_s,\quad \bar{d}^s(x_R,\teb)=E^{\dg s}-\sqrt{2}\teb v^{\dg s}-\teb^2 K^{\dg s},\\
&&\ml{Z}^A(x_L,\te)=Z^A+\sqrt{2}\te\zeta^A+\te^2 F^A,\quad \bar{\ml{Z}}_A(x_R,\teb)=Z^{\dg}_A-\sqrt{2}\teb\zeta^{\dg}_A-\teb^2 F^{\dg}_A,\\
&&\ml{W}_A(x_L,\te)=W_A+\sqrt{2}\te\omega_A+\te^2 G_A,\quad \bar{\ml{W}}^A(x_R,\teb)=W^{\dg A}-\sqrt{2}\teb\omega^{\dg A}-\teb^2 G^{\dg A},\\
&&\ml{V}=2i\theta\bar{\theta}\sigma(x)+2\theta\gamma^{\mu}\bar{\te}A_{\mu}(x)+\sqrt{2}i\te^2\bar{\te}\bar{\chi}(x)-\sqrt{2}i\bar{\te}^2\te\chi(x)+\te^2\bar{\te}^2D(x),\\
&&\ml{\hat{V}}=
2i\theta\bar{\theta}\hat{\sigma}(x)+2\theta\gamma^{\mu}\bar{\te}\hat{A}_{\mu}(x)+\sqrt{2}i\te^2\bar{\te}\hat{\bar{\chi}}(x)-\sqrt{2}i\bar{\te}^2\te\hat{\chi}(x)+\te^2\bar{\te}^2\hat{D}(x).
\eea
Notice that the expansions of the vector superfields are in Wess-Zumino gauge.

Following the treatment of deriving the component form of ABJM action \cite{Benna:2008}, we integrate out those auxiliary fields 
and then we find the total action becomes,
\bea
\ml{S}_{\ml{N}=2}&=&\int d^3x\,\, \left(\frac{k}{4\pi}\,\mbox{tr}\,\epsilon^{\mu\nu\lambda}(A_{\mu}\partial_{\nu}A_{\lambda}+\frac{2i}{3}A_{\mu}A_{\nu}A_{\lambda}-\hat{A}_{\mu}\partial_{\nu}\hat{A}_{\lambda}-\frac{2i}{3}\hat{A}_{\mu}\hat{A}_{\nu}\hat{A}_{\lambda})\right.\\\no
&&\left.-\mbox{tr} (\ml{D}^{\mu}Z)^{\dg}_A\ml{D}_{\mu}Z^A-\mbox{tr} (\ml{D}^{\mu}W)^{\dg A}\ml{D}_{\mu}W_A-\mbox{tr} (\ml{D}^{\mu}C)^{\dg}_s\ml{D}_{\mu}C^s-\mbox{tr} (\ml{D}^{\mu}B)^{\dg t}\ml{D}_{\mu}B_t\right.\\\no
&&\left.-\mbox{tr} (\ml{D}^{\mu}E)^{\dg s}\ml{D}_{\mu}E_s-\mbox{tr} (\ml{D}^{\mu}A)^{\dg}_t\ml{D}_{\mu}A^t-i\,\mbox{tr}\, \zeta^{\dg}_A\ml{D}\!\!\!\!/ \,\zeta^A-i\,\mbox{tr}\,\omega^{\dg A}\ml{D}\!\!\!\!/ \,\omega_A-i\,\mbox{tr}\,\tau^{\dg}_s\ml{D}\!\!\!\!/ \,\tau^s\right.\\\no
&&\left.-i\,\mbox{tr}\,\eta^{\dg t}\ml{D}\!\!\!\!/ \,\eta_t-i\,\mbox{tr}\,v^{\dg s}\ml{D}\!\!\!\!/ \,v_s-i\,\mbox{tr}\,\kappa^{\dg}_t\ml{D}\!\!\!\!/ \,\kappa^t-V_F^{bos}-V_F^{ferm}-V_D^{bos}-V_D^{ferm}\right),
\eea
where the covariant derivatives are defined as,
\bea
&\ml{D}_{\mu}{\Phi}^{A}=\partial_{\mu}{\Phi}^{A}+i A_{\mu}{\Phi}^{A}-i{\Phi}^{A}\hat{A}_{\mu},\quad \mbox{for} \quad \Phi^A\in (N_c,\bar{N_c}),\\
&\ml{D}_{\mu}\phi^s=\partial_{\mu}\phi^s+i A_{\mu} \phi^s,\quad \mbox{for} \quad \phi^s\in (N_c,1),\\
&\ml{D}_{\mu}\hat{\phi}_t=\partial_{\mu}\hat{\phi}_t+i \hat{A}_{\mu} \hat{\phi}_t,\quad \mbox{for} \quad \hat{\phi}_t\in (1,N_c).
\eea
We put the lengthy expressions of the potential terms in  appendix~\ref{appendixa} together with the on-shell values of auxiliary fields.

\subsection{The action in $\ml{N}=3$ component field formulation}
In order to obtain a manifestly $SU(2)_R$ invariant theory, we combine the component fields into the following doublet form
\bea
  &&X^{aA}=\left(
\ba{ll}
Z^A\\W^{\dg A}
\ea
\right),
\quad
 X^{\dg}_{aA}=\left(
\ba{ll}
Z^{\dg}_A\\W_A
\ea
\right),
\quad
\xi^{aA}=\left(
\ba{ll}
\omega^{\dg A}e^{{i\pi}/{4}}\\\zeta^Ae^{-{i\pi}/{4}}
\ea
\right),
\quad
\xi^{\dg}_{aA}=\left(
\ba{ll}
\omega_Ae^{-{i\pi}/{4}}\\\zeta^{\dg}_Ae^{{i\pi}/{4}}
\ea
\right),\\
&&Y^{as}=\left(
\ba{ll}
C^s\\E^{\dg s}
\ea
\right),
\quad
 Y^{\dg}_{as}=\left(
\ba{ll}
C^{\dg}_s\\E_s
\ea
\right),
\quad
\psi^{as}=\left(
\ba{ll}
v^{\dg s}e^{{i\pi}/{4}}\\ \tau^se^{-{i\pi}/{4}}
\ea
\right),
\quad
\psi^{\dg}_{as}=\left(
\ba{ll}
v_se^{-{i\pi}/{4}}\\\tau^{\dg}_se^{{i\pi}/{4}}
\ea
\right),\\
 &&M^{at}=\left(
\ba{ll}
A^t\\B^{\dg t}
\ea
\right),
\quad
 M^{\dg}_{at}=\left(
\ba{ll}
A^{\dg}_t\\B_t
\ea
\right),
\quad
\theta^{at}=\left(
\ba{ll}
\eta^{\dg t}e^{{i\pi}/{4}}\\ \kappa^te^{-{i\pi}/{4}}
\ea
\right),
\quad
\theta^{\dg}_{at}=\left(
\ba{ll}
\eta_te^{-{i\pi}/{4}}\\ \kappa^{\dg}_te^{{i\pi}/{4}}
\ea
\right),
\eea
where the explicit $SU(2)_R$ R-symmetry index $a$ is raised and lowered by the anti-symmetric tensor $\ep^{ab}$ and $\ep_{ab}$ with $\ep^{12}=-\ep_{12}=1$.\footnote{In the following we will also use $i, j, \cdots$ as R-symmetry indices.}
\par In light of the work in \cite{Benna:2009} where a $\ml{N}=3$ Chern-Simons Yang-Mills theory was given, we re-write the above action into a manifestly $SU(2)_R$ invariant form in terms of these new fields as
\bea
\ml{S}_{\ml{N}=3}&=&\int d^3x \, \mbox{tr}\left[\frac{k}{4\pi} \ep^{\mu\nu\lambda}\left(A_{\mu}\partial_{\nu}A_{\lambda}+\frac{2i}{3}A_{\mu}A_{\nu}A_{\lambda}\right)-
\frac{k}{4\pi} \ep^{\mu\nu\lambda}\left(\hat{A}_{\mu}\partial_{\nu}\hat{A}_{\lambda}+\frac{2i}{3}\hat{A}_{\mu}\hat{A}_{\nu}\hat{A}_{\lambda}\right)\right.\\\no
&&\left.-\mathcal{D}_{\mu}X^{\dg}_{aA}\mathcal{D}^{\mu}X^{aA}-\mathcal{D}_{\mu}Y^{\dg}_a\mathcal{D}^{\mu}Y^a-\mathcal{D}_{\mu}M^{\dg}_a\mathcal{D}^{\mu}M^a
+i\xi^{\dg}_{aA}\ml{D}\!\!\!\!/ \,\xi^{aA}+i\psi^{\dg}_a\ml{D}\!\!\!\!/ \,\psi^a+i\theta^{\dg}_a\ml{D}\!\!\!\!/ \,\theta^a\right.\\\no
&&\left.-V_{ferm}^{\ml{N}=3}-V^{\ml{N}=3}_{bos}\right.\bigg]
\eea
with the fermionic part of the potential\footnote{Our convention for symmetrization is $f_{(ab)}=\frac12(f_{ab}+f_{ba})$ and $f_{(a|B|b)}=\frac12(f_{aBb}+f_{bBa})$.}
\bea
&&-V_{ferm}^{\ml{N}=3}\\\no
&=& -\frac{2\pi i}{k}\ep_{bc}\ep_{ad}\left(\xi^{aA}X^{\dg b}_A-X^{bA}\xi^{\dg a}_A+\psi^aY^{\dg b}-Y^b\psi^{\dg a}\right)\left(\xi^{cB}X^{\dg d}_B-X^{dB}\xi^{\dg c}_B+\psi^cY^{\dg d}-Y^d\psi^{\dg c}\right)\\\no
&&\left.+\frac{2\pi i}{k}\ep_{bc}\ep_{ad}\left(-\xi^{\dg a}_AX^{bA}+X^{\dg b}_A\xi^{aA}-\te^{\dg a}M^b+M^{\dg b}\te^a\right)\left(-\xi^{\dg c}_BX^{dB}+X^{\dg d}_B\xi^{cB}-\te^{\dg c}M^d+M^{\dg d}\te^c\right)\right.\\\no
&&\left.+\frac{4\pi i}{k}\ep_{ac}\left(\xi^{aA}\xi^{\dg}_{bA}+\psi^a\psi^{\dg}_b\right)\left(X^{(c|B|}{X}^{\dg b)}_B+Y^{(c}Y^{\dg b)}\right)\right.\\\no
&&\left.-\frac{4\pi i}{k}\ep_{ac}\left(\xi^{\dg}_{bA}\xi^{aA}+\te^{\dg}_b\te^a\right)\left({X}^{\dg(c}_BX^{b)B}+{M}^{\dg(c}M^{b)}\right)
\right.,
\eea
and the bosonic part, which is first given in \cite{Hohenegger:2009}\footnote{In fact, there is a mistake in eq.(A.4) of the paper \cite{Hohenegger:2009}: the second and the fourth terms should be corrected as $\bar{q}^a_2q^2_a\bar{q}^b_2q^2_b\bar{q}^c_2q^2_c$ and $-4\bar{q}^a_2q^2_b\bar{q}^c_2q^2_a\bar{q}^b_2q^2_c$ , respectively.},
\bea
&&\no-V^{\ml{N}=3}_{bos}\\\no
&=&\frac{4\pi^2}{3k^2}\left[
Y^aY^{\dg}_aY^bY^{\dg}_bY^cY^{\dg}_c+M^{\dg}_aM^aM^{\dg}_bM^bM^{\dg}_cM^c-4Y^aY^{\dg}_bY^cY^{\dg}_aY^bY^{\dg}_c-4M^{\dg}_aM^bM^{\dg}_cM^aM^{\dg}_bM^c
\right.\\\no
&&\left.+X^{aA}X^{\dg}_{aA}X^{bB}X^{\dg}_{bB}X^{cM}X^{\dg}_{cM}+X^{\dg}_{aA}X^{aA}X^{\dg}_{bB}X^{bB}X^{\dg}_{cM}X^{cM}+4X^{aA}X^{\dg}_{bB}X^{cM}X^{\dg}_{aA}X^{bB}X^{\dg}_{cM}
\right.\\\no
&&\left.
-6X^{aA}X^{\dg}_{bB}X^{bB}X^{\dg}_{aA}X^{cM}X^{\dg}_{cM}+3X^{aA}X^{\dg}_{aA}X^{bB}X^{\dg}_{bB}Y^cY^{\dg}_c+3X^{\dg}_{aA}X^{aA}X^{\dg}_{bB}X^{bB}M^{\dg}_cM^c\right.\\\no
&&\left.
-6X^{aA}X^{\dg}_{bB}X^{bB}X^{\dg}_{aA}Y^cY^{\dg}_c-6X^{\dg}_{aA}X^{bB}X^{\dg}_{bB}X^{aA}M^{\dg}_cM^c+9X^{aA}X^{\dg}_{aA}Y^bY^{\dg}_bY^cY^{\dg}_c\right.\\
&&\left.
+9X^{\dg}_{aA}X^{aA}M^{\dg}_bM^bM^{\dg}_cM^c-6X^{aA}X^{\dg}_{aA}Y^bY^{\dg}_cY^cY^{\dg}_b-6X^{\dg}_{aA}X^{aA}M^{\dg}_bM^cM^{\dg}_cM^b\right.\\\no
&&\left.
-6X^{aA}X^{\dg}_{bA}Y^bY^{\dg}_aY^cY^{\dg}_c-6X^{\dg}_{aA}X^{bA}M^{\dg}_bM^aM^{\dg}_cM^c+6X^{aA}X^{\dg}_{bA}Y^bY^{\dg}_cY^cY^{\dg}_a\right.\\\no
&&\left.
+6X^{\dg}_{aA}X^{bA}M^{\dg}_bM^cM^{\dg}_cM^a-6X^{aA}X^{\dg}_{bA}Y^cY^{\dg}_aY^bY^{\dg}_c-6X^{\dg}_{aA}X^{bA}M^{\dg}_cM^aM^{\dg}_bM^c\right.\\\no
&&\left.
-6X^{aA}X^{\dg}_{bA}Y^cY^{\dg}_cY^bY^{\dg}_a-6X^{\dg}_{aA}X^{bA}M^{\dg}_cM^cM^{\dg}_bM^a-6X^{\dg}_{aA}Y^bY^{\dg}_bX^{aA}M^{\dg}_cM^c\right.\\\no
&&\left.
+12X^{\dg}_{aA}Y^bY^{\dg}_cX^{aA}M^{\dg}_bM^c+12\ep_{AB}\ep^{MN}X^{cA}X^{\dg}_{bM}X^{aB}X^{\dg}_{cN}Y^bY^{\dg}_a\right.\\\no
&&\left.+12\ep^{AB}\ep_{MN}X^{\dg}_{cA}X^{bM}X^{\dg}_{aB}X^{cN}M^{\dg}_bM^a\right],
\eea
where flavor indices are suppressed.
\par Thus, we demonstrate the enhancement of the R-symmetry to $SU(2)_R$ by the explicit construction of the action. Besides the $SU(2)_R$ symmetry, the theory also has $SU(2)_D$ symmetry acting on the $A$ index of $X^{aA}$. The above action is the starting point of our perturbative calculations.

\section{Two-loop perturbative calculations and the Hamiltonian}\label{pert}
In this section, we will compute the ADM of gauge invariant composite operators. We will perform the calculations in the 't Hooft limt with  $N_c\rightarrow \infty$, $k\rightarrow \infty$ while $\lambda\equiv N_c/k, N_{f_1}, N_{f_2}$ fixed.\footnote{We will further set $N_{f_1}=N_{f_2}=1$ without loss of generality and then neglect the flavor indices for simplicity.} Since the ADM of single trace operators is the same as the one in the ABJM theory, we only need to consider the mesonic operators.
We focus on two types of mesonic operators,\footnote{There exist two other types of composite operators sharing essentially the same structures as those considered in the main text, namely, $MX^{\dg}X\cdots X^{\dg}XM^{\dg}$ and $MX^{\dg}X\cdots X^{\dg}XX^{\dg}Y$, and we will not repeat the similar analysis here.}
\bea
\hat{O}&=&Y^{\dg}_iX^{i_1A_1}X^{\dg}_{i_2A_2}\cdots X^{i_{2L-1}A_{2L-1}}X^{\dg}_{i_{2L}A_{2L}}Y^{i'},\\
\hat{O}'&=&Y^{\dg}_iX^{i_1A_1}X^{\dg}_{i_2A_2}\cdots X^{i_{2L-1}A_{2L-1}}X^{\dg}_{i_{2L}A_{2L}}X^{i_{2L+1}A_{2L+1}}M^{\dg}_{i'},
\eea
where $L>2$ and the contraction of the color indices is implied. We note that these composite operators are built up without trace operations since they are bounded on both sides by (anti-) fundamental matters. Our aim is to extract  the ADM from the two point correlation function through  2-loop Feynman diagram computations. The calculations concerning only the bi-fundamental fields in the bulk are the same as those for the single trace operator $\mbox{tr}(XX^{\dg}\cdots XX^{\dg})$ in ABJM theory and have been carried out carefully in \cite{Minahan:2008,Bak:2008}\footnote{In marginally deformed ABJM theories,  similar calculations have been performed in \cite{He:2013hxd, Chen:2016geo}.}. Here we will concentrate on the boundary part and show the details of the derivation of ADM of $\hat{O}$. For the operator $\hat{O}'$, the whole procedure is identical and we will give the result directly in the end.
\subsection{Boundary three-site scalar interactions}
\begin{figure}[H]
\begin{center}
\includegraphics[scale=0.5]{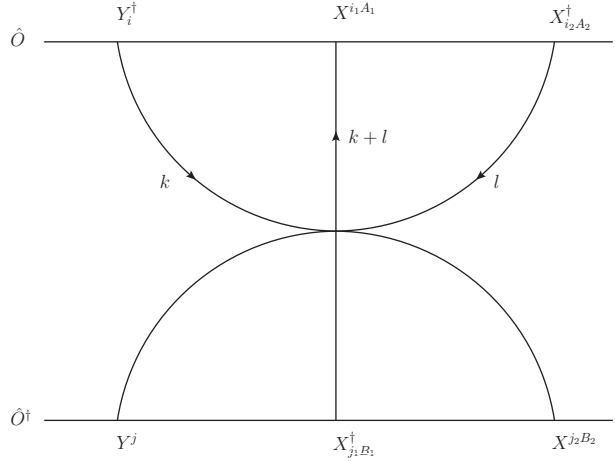}
\end{center}
\caption{\sl The contribution of six scalar interaction vertex on the left boundary}
\label{sixscalar}
\end{figure}

First we compute the contribution of the six-point vertex on the left boundary shown in Fig.\ref{sixscalar}.
The relevant interaction terms in the $\ml{N}=3$ Lagrangian are:
\bea
&&V_B^1=\frac{4\pi^2}{k^2}X^{aA}X^{\dg}_{aA}X^{bB}X^{\dg}_{bB}Y^cY^{\dg}_c,\\
&&V_B^2=-\frac{8\pi^2}{k^2}X^{aA}X^{\dg}_{bB}X^{bB}X^{\dg}_{aA}Y^cY^{\dg}_c,\\
&&V_B^3=\frac{16\pi^2}{k^2}\ep_{AB}\ep^{MN}X^{cA}X^{\dg}_{bM}X^{aB}X^{\dg}_{cN}Y^bY^{\dg}_a.
\eea
Let us analyze these interaction vertices separately and mainly focus on the flavor structure as follows:
\begin{itemize}
  \item $V_B^1$:
  \bea
  \de^c_i\de^{i_1}_b\de^{A_1}_B\de^b_{i_2}\de^B_{A_2}\de^{j_2}_a\de^{B_2}_A\de^a_{j_1}\de^A_{B_1}\de^j_c=\de^j_i\de^{i_1}_{i_2}\de^{j_2}_{j_1}\cdot
  \de^{A_1}_{A_2}\de^{B_2}_{B_1}.
  \eea
  \item $V_B^2$:
  \bea
  \de^j_c\de^c_i\de^{i_1}_a\de^{A_1}_A\de^b_{i_2}\de^B_{A_2}\de^{j_2}_b\de^{B_2}_B\de^a_{j_1}\de^A_{B_1}=\de^j_i\de^{i_1}_{j_1}\de^{j_2}_{i_2}\cdot
  \de^{A_1}_{B_1}\de^{B_2}_{A_2}.
  \eea
  \item $V_B^3$:
  \bea
  \de^j_a\de^b_i\de^{i_1}_c\de^{A_1}_N\de^a_{i_2}\de^B_{A_2}\de^{j_2}_b\de^{B_2}_M\de^c_{j_1}\de^A_{B_1}\ep_{AB}\ep^{MN}=-\de^{j_2}_i\de^j_{i_2}\de^{i_1}_{j_1}
  \cdot\left(\de^{B_2}_{B_1}\de^{A_1}_{A_2}-\de^{A_1}_{B_1}\de^{B_2}_{A_2}\right).
  \eea
\end{itemize}
We will use dimensional regularization to isolate the divergence and set $d=3-\epsilon$ with the relation: $\ep^{-1}=\log \Lambda^2$ where $\Lambda$ is the momentum space cutoff. The two-loop integral in Fig.\ref{sixscalar} reads
\bea
(-i)^3[i]\cdot i^2\int\frac{d^dk}{(2\pi)^d}\frac{d^dl}{(2\pi)^d}\frac{1}{(k+l)^2}\frac{1}{k^2}\frac{1}{l^2}=\frac{1}{16\pi^2}\log\Lambda,
\eea
where the factor $[i]$ comes from the six-point vertex and $(-i)^3$ comes from the scalar propagator. The rest part of the above formula is a loop integral evaluated in Euclidean space and the factor $i^2$ accounts for the Wick rotation. There is also a factor $N_c^2$ from the contraction of color indices. Putting these together and noting that the contribution to the operator renormalization  is negative of the quantum correction, we find the left boundary three-site scalar interaction gives
\bea
\left({\ml{H}}^B_l\right)^{j, i_1A_1, j_2B_2}_{i, j_1B_1, i_2A_2}=-\frac{\lambda^2}{4}\left[\left(\de^j_i\de^{i_1}_{i_2}\de^{j_2}_{j_1}-4\de^{j_2}_i\de^j_{i_2}\de^{i_1}_{j_1}\right)\de^{A_1}_{A_2}\de^{B_2}_{B_1}
-2\left(\de^j_i\de^{i_1}_{j_1}\de^{j_2}_{i_2}-2\de^{j_2}_i\de^{i_1}_{j_1}\de^j_{i_2}\right)\de^{A_1}_{B_1}\de^{B_2}_{A_2}\right].
\eea
The contribution from the right boundary can be obtained simply by some replacements of indices and we get
\bea
\left({\ml{H}}^B_r\right)^{i_{2L-1}A_{2L-1}, j_{2L}B_{2L}, i'}_{j_{2L-1}B_{2L-1}, i_{2L}A_{2L}, j'}=-\frac{\lambda^2}{4}\left[\left(\de^{i'}_{j'}\de^{j_{2L}}_{j_{2L-1}}\de^{i_{2L-1}}_{i_{2L}}-4\de^{i_{2L-1}}_{j'}\de^{i'}_{j_{2L-1}}\de^{j_{2L}}_{i_{2L}}\right)
\de^{B_{2L}}_{B_{2L-1}}\de^{A_{2L-1}}_{A_{2L}}\right.\\\no
\left.-2\left(\de^{i'}_{j'}\de^{j_{2L}}_{i_{2L}}\de^{i_{2L-1}}_{j_{2L-1}}-2\de^{i_{2L-1}}_{j'}\de^{j_{2L}}_{i_{2L}}\de^{i'}_{j_{2L-1}}\right)\de^{B_{2L}}_{A_{2L}}\de^{A_{2L-1}}_{B_{2L-1}}\right].
\eea
\subsection{Boundary two-site Yukawa type interactions}
\begin{figure}[H]
\begin{center}
\includegraphics[scale=0.5]{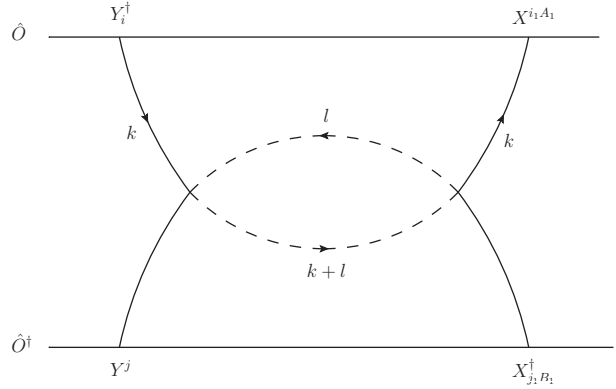}
\end{center}
\caption{\sl The contribution of Yukawa interactions on the left boundary}
\label{fermloop}
\end{figure}

The Feynman diagram of the boundary two-site contribution consists of two Yukawa type vertices and a fermion loop depicted in Fig.\ref{fermloop}.
The involved interaction vertices are listed below:
\bea
V_F^1&=&\frac{4\pi i}{k} X^{aB}X^{\dg b}_A\xi^A_a\xi^{\dg}_{bB},\\
V_F^2&=&-\frac{2\pi i}{k}X^{aB}X^{\dg b}_B\xi^A_a\xi^{\dg}_{bA},\\
V_F^3&=&-\frac{2\pi i}{k}X^{aB}X^{\dg b}_B\xi^A_b\xi^{\dg}_{aA},\\
\tilde{V}_F^1&=&-\frac{2\pi i}{k}Y^aY^{\dg b}\xi^A_a\xi^{\dg}_{bA},\\
\tilde{V}_F^2&=&-\frac{2\pi i}{k}Y^aY^{\dg b}\xi^A_b\xi^{\dg}_{aA}.
\eea
We have ignored the diagrams whose internal fermions belong to the fundamental flavors because such diagrams will be generically suppressed by a factor of $N_f/N_c$ in the 't Hooft limit. Now let us investigate the flavor structure arising from all possible combinations of the above vertices.
\begin{itemize}
  \item $V_F^1\otimes\tilde{V}_F^1$:
  \bea
  \de^a_i\ep_{mb}\ep_{an}\ep^{ni_1}\de^m_{j_1}\ep^{bj}\cdot\de^Q_A\de^A_P\de^{A_1}_Q\de^P_{B_1}=\de^{i_1}_i\de^j_{j_1}\de^{A_1}_{B_1},
  \eea
  \item $V_F^1\otimes\tilde{V}_F^2$:
  \bea
  \de^a_i\ep_{ma}\ep_{bn}\ep^{ni_1}\de^m_{j_1}\ep^{bj}\cdot\de^Q_A\de^A_P\de^{A_1}_Q\de^P_{B_1}=\left(-\de^{i_1}_{j_1}\de^j_i+\de^j_{j_1}\de^{i_1}_i\right)\de^{A_1}_{B_1},
  \eea
  \item $V_F^2\otimes\tilde{V}_F^1$:
  \bea
  \de^a_i\ep_{mb}\ep_{an}\ep^{ni_1}\de^m_{j_1}\ep^{bj}\cdot\de^Q_A\de^A_Q\de^{A_1}_P\de^P_{B_1}=2\de^j_{j_1}\de^{i_1}_i\de^{A_1}_{B_1},
  \eea
  \item $V_F^2\otimes\tilde{V}_F^2$:
  \bea
  \de^a_i\ep_{ma}\ep_{bn}\ep^{ni_1}\de^m_{j_1}\ep^{bj}\cdot\de^Q_A\de^A_Q\de^{A_1}_P\de^P_{B_1}=2\left(-\de^{i_1}_{j_1}\de^j_i+\de^j_{j_1}\de^{i_1}_i\right)\de^{A_1}_{B_1},
  \eea
  \item $V_F^3\otimes\tilde{V}_F^1$:
  \bea
  \de^a_i\ep_{nb}\ep_{am}\ep^{ni_1}\de^m_{j_1}\ep^{bj}\cdot\de^Q_A\de^A_Q\de^{A_1}_P\de^P_{B_1}=2\left(-\de^j_i\de^{i_1}_{j_1}+\de^{i_1}_i\de^j_{j_1}\right)\de^{A_1}_{B_1},
  \eea
  \item $V_F^3\otimes\tilde{V}_F^2$:
  \bea
  \de^a_i\ep_{na}\ep_{bm}\ep^{ni_1}\de^m_{j_1}\ep^{bj}\cdot\de^Q_A\de^A_Q\de^{A_1}_P\de^P_{B_1}=2\de^{i_1}_i\de^j_{j_1}\de^{A_1}_{B_1}.
  \eea
\end{itemize}
The remaining loop integral is
\bea
\frac{2}{2!}(-i)^2(i)^2[i]^2\cdot (-1)i^2\int\frac{d^dk}{(2\pi)^d}\frac{d^dl}{(2\pi)^d}\frac{1}{(k^2)^2}\mbox{tr} \left(\frac{\gamma^{\mu}l_{\mu}}{l^2}\frac{\gamma^{\nu}(k+l)_{\nu}}{(k+l)^2}\right)=\frac{1}{16\pi^2}\log\Lambda,
\eea
where the factor ${2}/{2!}$ is from the coefficient of the second order expansion of the interaction terms and $(-i)^2,(i)^2,[i]^2$ come from the scalar and fermion propagators and the vertices respectively. The factor $(-1)$ is from fermion loop accounting for the Fermi-Dirac statistics. Gathering all these data, we find the final counter-term contributing to the dilatation operator is
\bea
\left({\ml{H}}^F_l\right)^{j, i_1A_1, j_2B_2}_{i, j_1B_1, i_2A_2}=\frac{\lambda^2}{2}\left(2\de^{i_1}_i\de^j_{j_1}-\de^j_i\de^{i_1}_{j_1}\right)\de^{j_2}_{i_2}\de^{A_1}_{B_1}\de^{B_2}_{A_2}.
\eea
For the right boundary, it is
\bea
\left({\ml{H}}^F_r\right)^{i_{2L-1}A_{2L-1}, j_{2L}B_{2L}, i'}_{j_{2L-1}B_{2L-1}, i_{2L}A_{2L}, j'}=\frac{\lambda^2}{2}\left(2\de^{j_{2L}}_{j'}\de^{i'}_{i_{2L}}-\de^{i'}_{j'}\de^{j_{2L}}_{i_{2L}}\right)\de^{i_{2L-1}}_{j_{2L-1}}
\de^{B_{2L}}_{A_{2L}}\de^{A_{2L-1}}_{B_{2L-1}}.
\eea
\subsection{The two-loop Hamiltonian}
There is another two-site diagram concerning the exchange interaction of gauge bosons, however, this Feynman diagram can only give constant  contribution since the gauge propagators do not carry flavor indices. As for the various one-site diagrams representing the wave function renormalization, it is easy to see that they also lead to constant pieces.
Note also that the two two-bulk-site trace operators in $\ml{H}^B_l$ and $\ml{H}^B_r$ are canceled by the bulk two-site interactions. And this cancelation makes the whole bulk Hamiltonian to be in the same form as the one derived from the single trace operator in ABJM theory. Finally, the two-loop Hamiltonian associated with the ADM  of the composite operator $\hat{O}$ reads
\bea
{\ml{H}}={\ml{H}}_l+{\ml{H}}_r+{\ml{H}}_{bulk}+\al\mathbb{I},\label{H1}
\eea
with
\bea
\left({\ml{H}}_l\right)^{j, i_1A_1, j_2B_2}_{i, j_1B_1, i_2A_2}=\lambda^2\left[\left(\de^{A_1}_{A_2}\de^{B_2}_{B_1}-\de^{A_1}_{B_1}\de^{B_2}_{A_2}\right)\cdot\de^{j_2}_i\de^{i_1}_{j_1}\de^j_{i_2}
+\de^{A_1}_{B_1}\de^{B_2}_{A_2}\cdot\de^{i_1}_i\de^j_{j_1}\de^{j_2}_{i_2}\right],
\eea
\bea
\left({\ml{H}}_r\right)^{i_{2L-1}A_{2L-1}, j_{2L}B_{2L}, i'}_{j_{2L-1}B_{2L-1}, i_{2L}A_{2L}, j'}
=\lambda^2\left[\left(\de^{B_{2L}}_{B_{2L-1}}\de^{A_{2L-1}}_{A_{2L}}-\de^{B_{2L}}_{A_{2L}}\de^{A_{2L-1}}_{B_{2L-1}}\right)\cdot\de^{i_{2L-1}}_{j'}\de^{j_{2L}}_{i_{2L}}\de^{i'}_{j_{2L-1}}
\right.\\\no
\left.+\de^{B_{2L}}_{A_{2L}}\de^{A_{2L-1}}_{B_{2L-1}}\cdot\de^{i_{2L-1}}_{j_{2L-1}}\de^{i'}_{i_{2L}}\de^{j_{2L}}_{j'}\right],
\eea
\bea
{\ml{H}}_{bulk}=\lambda^2\sum_{l=1}^{2L-2}\left(\mathbb{I}_{l,l+1}-\mathbb{P}_{l,l+2}+\frac{1}{2}\mathbb{P}_{l,l+2}\mathbb{K}_{l,l+1}+\frac{1}{2}\mathbb{K}_{l,l+1}\mathbb{P}_{l,l+2}\right),
\eea
where the basic operators $\mathbb{I}$, $\mathbb{P}$ and $\mathbb{K}$ are defined as
\bea
\left(\mathbb{I}_{l,l+1}\right)^{iA,\,j'B'}_{jB,\,i'A'}=\de^i_j\de^{j'}_{i'}\de^A_B\de^{B'}_{A'},\quad \left(\mathbb{P}_{l,l+2}\right)^{iA,\,i'A'}_{jB,\,j'B'}=\de^i_{j'}\de^{i'}_j\de^A_{B'}\de^{A'}_B,\quad
\left(\mathbb{K}_{l,l+1}\right)^{iA,\,j'B'}_{jB,\,i'A'}=\de^i_{i'}\de^{j'}_j\de^A_{A'}\de^{B'}_B,
\eea
and the exact value of the coefficient $\al$ will be determined later by the BPS condition of the corresponding vacuum state. For operator $\hat{O}'$, the two-loop Hamiltonian is
\bea
{\ml{H}}'={\ml{H}}'_l+{\ml{H}}'_r+{\ml{H}}'_{bulk}+\al' \mathbb{I},
\eea
where
\bea
{\ml{H}}'_l={\ml{H}}_l,
\eea
\bea
\left({\ml{H}}'_r\right)^{j_{2L}B_{2L}, i_{2L+1}A_{2L+1}, j'}_{i_{2L}A_{2L}, j_{2L+1}B_{2L+1}, i'}=\lambda^2\left[\left(\de^{A_{2L+1}}_{A_{2L}}\de^{B_{2L}}_{B_{2L+1}}-\de^{A_{2L+1}}_{B_{2L+1}}\de^{B_{2L}}_{A_{2L}}\right)\cdot\de^{j_{2L}}_{i'}\de^{i_{2L+1}}_{j_{2L+1}}\de^{j'}_{i_{2L}}
\right.\\\no
\left.+\de^{A_{2L+1}}_{B_{2L+1}}\de^{B_{2L}}_{A_{2L}}\cdot\de^{i_{2L+1}}_{i'}\de^{j'}_{j_{2L+1}}\de^{j_{2L}}_{i_{2L}}\right],
\eea
\bea
{\ml{H}}'_{bulk}=\lambda^2\sum_{l=1}^{2L-1}\left(\mathbb{I}_{l,l+1}-\mathbb{P}_{l,l+2}+\frac{1}{2}\mathbb{P}_{l,l+2}\mathbb{K}_{l,l+1}+\frac{1}{2}\mathbb{K}_{l,l+1}\mathbb{P}_{l,l+2}\right).
\eea
We would like to mention some features of the boundary interaction. It breaks the $SU(4)_R$ symmetry of the bulk interaction into $SU(2)_R\times SU(2)_D$.
It includes both nearest and next-to-nearest neighbour interactions, especially the two-site trace operators \footnote{These involve the boundary site and the nearest bulk site.} which do not appear in the bulk interaction.
\section{Integrability from coordinate Bethe ansatz}\label{cba}
In this section, we will prove the integrability of  flavored ABJM model by showing that the boundary reflection matrices satisfy the reflection equations. These reflection matrices are obtained by the concrete constructions of Bethe ansatz solutions of the Hamiltonian. We begin with the composite operator $\hat{O}$ which naturally corresponds to an open spin chain state and the vacuum or the Bethe reference state is chosen to be
\bea
|\Omega\rangle=|Y^{\dg}_2X^{11}X^{\dg}_{22}\cdots X^{11}X^{\dg}_{22}Y^1\rangle.
\eea
For the case of single impurities, the Hamiltonian in eq.(\ref{H1}) reduces to
\bea
\ml{H}=\ml{H}_l+\ml{H}_r+\al \mathbb{I}+\lambda^2\sum^{2L-2}_{l=1}\left(\mathbb{I}_{l,l+1}-\mathbb{P}_{l,l+2}\right).
\eea
In appendix~\ref{susy}, we will demonstrate that the vacuum is  a BPS state, so its scaling dimension receives no quantum corrections. This determines $\al$ to be $2\lambda^2$.
We now use a conventional way to label the bulk fields as A and B types as follows,
\bea
X^{11}=A_1,\quad X^{12}=A_2,\quad X^{21}=B^{\dg}_1,\quad X^{22}=B^{\dg}_2.
\eea
There are three types of one-particle excitations,
\bea
\mbox{bulk A type}: \quad Y^{\dg}_2(A_1B_2)\cdots(A_2B_2)\cdots(A_1B_2)Y^1,&&\\
Y^{\dg}_2(A_1B_2)\cdots(B^{\dg}_1B_2)\cdots(A_1B_2)Y^1,&&\\
\mbox{bulk B type}: \quad Y^{\dg}_2(A_1B_2)\cdots(A_1A^{\dg}_2)\cdots(A_1B_2)Y^1,&&\\
Y^{\dg}_2(A_1B_2)\cdots(A_1B_1)\cdots(A_1B_2)Y^1,&&\\
\mbox{boundary\quad}: \quad Y^{\dg}_1(A_1B_2)\cdots(A_1B_2)\cdots(A_1B_2)Y^1,&&\\
Y^{\dg}_2(A_1B_2)\cdots(A_1B_2)\cdots(A_1B_2)Y^2.&&
\eea
After scattering at the boundary, these pseudo-particles will transform into each other. Under the action of $\ml{H}_l$,
\bea
&&\ml{H}_l|1\rangle_{A_2}=\lambda^2|1\rangle_{B_1},\\
&&\ml{H}_l|1\rangle_{B^{\dg}_1}=\lambda^2|l\rangle_{Y^{\dg}_1},\\
&&\ml{H}_l|1\rangle_{A_2^{\dg}}=-\lambda^2|l\rangle_{Y^{\dg}_1},\\
&&\ml{H}_l|1\rangle_{B_1}=\lambda^2|1\rangle_{A_2},\\
&&\ml{H}_l|l\rangle_{Y^{\dg}_1}=-\lambda^2|1\rangle_{A_2^{\dg}}+\lambda^2|l\rangle_{Y^{\dg}_1}+\lambda^2|1\rangle_{B^{\dg}_1},\\
&&\ml{H}_l|x\rangle=-\lambda^2|x\rangle, \quad x\neq 1,
\eea
and under the action of $\ml{H}_r$,
\bea
&&\ml{H}_r|L\rangle_{A_2}=\lambda^2|L\rangle_{B_1},\\
&&\ml{H}_r|L\rangle_{B_1^{\dg}}=-\lambda^2|r\rangle_{Y^2},\\
&&\ml{H}_r|L\rangle_{A_2^{\dg}}=\lambda^2|r\rangle_{Y^2},\\
&&\ml{H}_r|L\rangle_{B_1}=\lambda^2|L\rangle_{A_2},\\
&&\ml{H}_r|r\rangle_{Y^2}=-\lambda^2|L\rangle_{B_1^{\dg}}+\lambda^2|r\rangle_{Y^2}+\lambda^2|L\rangle_{A_2^{\dg}},\\
&&\ml{H}_r|x\rangle=-\lambda^2|x\rangle, \quad x\neq L,
\eea
where the spin chain is symbolised as $|l(1)(2)\cdots(x)\cdots(L)r\rangle$ with every site $(x)$ containing two fields. We use  the excitation with its position  to label  the state and  use $|x\rangle$ without subscript to denote any of $|x\rangle_{A_2}, |x\rangle_{B_1}, |x\rangle_{A_2^\dg}, |x\rangle_{B_1^\dg}$. Here we see the novelty of our model where different states can mix and nontrivial boundary reflections will appear unlike those parallel studies in SYM with fundamental matters \cite{DeWolfe:2004, Erler:2005nr}. Then we find that only the superposition of several different one-particle spin wave functions can be constructed as an eigenstate of the Hamiltonian and we can extract the boundary reflection matrix by solving the corresponding eigenvalue equations.
\subsection{Two-particle mixed sector}
We consider the superposition of the spin wave functions of particles $A_2$ and $B_1$ as follows,
\bea
|\psi_1(k)\rangle=\sum_{x=1}^{L}\left(f(x)|x\rangle_{A_2}+g(x)|x\rangle_{B_1}\right),
\eea
where the Bethe ansatz for the wave functions are
\bea
f(x)=Fe^{ikx}+\tilde{F}e^{-ikx},\label{BA1}\\
g(x)=Ge^{ikx}+\tilde{G}e^{-ikx}.\label{BA2}
\eea
Using our Hamiltonian, we find that
\bea
\ml{H}\sum_{x=1}^{L}f(x)|x\rangle_{A_2}&=&\lambda^2f(1)|1\rangle_{B_1}+\lambda^2f(L)|L\rangle_{B_1}
+2\lambda^2f(1)|1\rangle_{A_2}+2\lambda^2f(L)|L\rangle_{A_2}\\\no
&&-\lambda^2f(1)|2\rangle_{A_2}-\lambda^2f(L)|L-1\rangle_{A_2}\\\no
&&+\lambda^2\sum_{x=2}^{L-1}f(x)\left(2|x\rangle_{A_2}-|x-1\rangle_{A_2}-|x+1\rangle_{A_2}\right),
\eea
and
\bea
\ml{H}\sum_{x=1}^{L}g(x)|x\rangle_{B_1}&=&\lambda^2g(1)|1\rangle_{A_2}+\lambda^2g(L)|L\rangle_{A_2}
+2\lambda^2g(1)|1\rangle_{B_1}+2\lambda^2g(L)|L\rangle_{B_1}\\\no
&&-\lambda^2g(1)|2\rangle_{B_1}-\lambda^2g(L)|L-1\rangle_{B_1}\\\no
&&+\lambda^2\sum_{x=2}^{L-1}g(x)\left(2|x\rangle_{B_1}-|x-1\rangle_{B_1}-|x+1\rangle_{B_1}\right),
\eea
The eigenvalue equation $\ml{H}|\psi_1\rangle=E(k)|\psi_1\rangle$ gives:
\begin{itemize}
  \item The bulk part ($x\neq 1,L$),
  \bea
  &2\lambda^2f(x)-\lambda^2f(x+1)-\lambda^2f(x-1)=E f(x),\label{EVE1}\\
  &2\lambda^2g(x)-\lambda^2g(x+1)-\lambda^2g(x-1)=E g(x).\label{EVE2}
  \eea
  Substituting eq.(\ref{BA1}) and eq.(\ref{BA2}) into the above equations, we have the following dispersion relation for the proposed spin wave,
  \bea
  E(k)=2\lambda^2-2\lambda^2\cos k.
  \eea
  \item The left boundary part,
  \bea
  \lambda^2f(1)+2\lambda^2g(1)-\lambda^2g(2)=Eg(1),\\
  2\lambda^2f(1)-\lambda^2f(2)+\lambda^2g(1)=Ef(1).
  \eea
  Using eqs.(\ref{EVE1}) and (\ref{EVE2}), the above coupled equations become
  \bea
  f(1)=-g(0),\\
  g(1)=-f(0).
  \eea
  By the plane wave expansions of eq.(\ref{BA1}) and eq.(\ref{BA2}), we find the relations
  \bea
  &Fe^{ik}+\tilde{F}e^{-ik}+G+\tilde{G}=0,\label{eq1}\\
  &Ge^{ik}+\tilde{G}e^{-ik}+{F}+{\tilde{F}}=0.\label{eq2}
  \eea
  The solution is
  \bea
  F=-e^{-ik}\tilde{G}, \quad G=-e^{-ik}\tilde{F}. \label{FG}
  \eea
  We define the left reflection matrix $K_l(k)$ in this sector by
  \be
 \left(\begin{array}{c}F\\
  G\end{array}\right)\equiv K_l(k)\left(\begin{array}{c}\tilde{F}\\
  \tilde{G}\end{array}\right).\ee
So from the above solution, we have,
  \bea K_l(k)=\left(\begin{array}{cc}0&-e^{-ik}\\
  -e^{-ik}&0\end{array}\right).
  \eea
  \item The right boundary part,
  \bea
  \lambda^2f(L)+2\lambda^2g(L)-\lambda^2g(L-1)=Eg(L),\\
  2\lambda^2f(L)-\lambda^2f(L-1)+\lambda^2g(L)=Ef(L),
  \eea
  which can be reduced to
  \bea
  f(L)+g(L+1)=0,\\
  g(L)+f(L+1)=0.
  \eea
  This gives \bea F e^{ikL}+\tilde{F}e^{-ikL}+Ge^{ik(L+1)}+\tilde{G}e^{-ik(L+1)}&=&0, \\
   G e^{ikL}+\tilde{G}e^{-ikL}+Fe^{ik(L+1)}+\tilde{F}e^{-ik(L+1)}&=&0.\eea
  Solving the above two equations, we get
  \bea F=-e^{-2ikL-ik}\tilde{G}, \quad \, G=-e^{-2ikL-ik}\tilde{F}. \label{FG2} \eea
  Following \cite{Li:2013vdh}, we define the right reflection matrix $K_r(k)$ in this sector by
\be\label{right}
 e^{2ikL}\left(\begin{array}{c}F\\
  G\end{array}\right)\equiv K_r(k)\left(\begin{array}{c}\tilde{F}\\
  \tilde{G}\end{array}\right).\ee
Then we get
  \bea K_r(k)=\left(\begin{array}{cc}0&-e^{-ik}\\
  -e^{-ik}&0\end{array}\right).
 \eea
  The consistency of eq.~(\ref{FG}) and (\ref{FG2}) gives 
  \bea
  e^{2ikL}=1,\quad k=\frac{n\pi}{L}, \quad n \in \mathbb{Z}.
  \eea
  This is the quantization conditions for our spin wave momenta $k$ as well as the Bethe equation for this two-particle mixed sector.
\end{itemize}
\subsection{Four-particle mixed sector}
Now we turn to another closed sector which consists of four excitations $B^{\dg}_1$, $A^{\dg}_2$, $Y^{\dg}_1$ and $Y^2$. The spin wave takes the form
\bea
|\psi_2(k)\rangle=\sum_{x=1}^L n(x)|x\rangle_{B^{\dg}_1}+\sum_{x=1}^{L}h(x)|x\rangle_{A_2^{\dg}}+\beta|l\rangle_{{Y^{\dg}_1}}+\gamma|r\rangle_{Y^2},
\eea
with
\bea
n(x)=Ne^{ikx}+\tilde{N}e^{-ikx},\label{BA3}\\
h(x)=He^{ikx}+\tilde{H}e^{-ikx}.\label{BA4}
\eea
The Hamiltonian acts on the above wave function as follows
\bea
\ml{H}\sum_{x=1}^{L}n(x)|x\rangle_{B^{\dg}_1}&=&\lambda^2 n(1)|l\rangle_{Y^{\dg}_1}-\lambda^2 n(L)|r\rangle_{Y^2}
+2\lambda^2 n(1)|1\rangle_{B^{\dg}_1}+2\lambda^2 n(L)|L\rangle_{B^{\dg}_1}\\\no
&&-\lambda^2 n(1)|2\rangle_{B^{\dg}_1}-\lambda^2 n(L)|L-1\rangle_{B^{\dg}_1}\\\no
&&+\lambda^2\sum_{x=2}^{L-1}n(x)\left(2|x\rangle_{B^{\dg}_1}-|x-1\rangle_{B^{\dg}_1}-|x+1\rangle_{B^{\dg}_1}\right),
\eea
\bea
\ml{H}\sum_{x=1}^{L}h(x)|x\rangle_{A^{\dg}_2}&=&-\lambda^2 h(1)|l\rangle_{Y^{\dg}_1}+\lambda^2 h(L)|r\rangle_{Y^2}
+2\lambda^2 h(1)|1\rangle_{A^{\dg}_2}+2\lambda^2 h(L)|L\rangle_{A^{\dg}_2}\\\no
&&-\lambda^2 h(1)|2\rangle_{A^{\dg}_2}-\lambda^2 h(L)|L-1\rangle_{A^{\dg}_2}\\\no
&&+\lambda^2\sum_{x=2}^{L-1}h(x)\left(2|x\rangle_{A^{\dg}_2}-|x-1\rangle_{A^{\dg}_2}-|x+1\rangle_{A^{\dg}_2}\right),
\eea
and
\bea
&\ml{H}|l\rangle_{Y^{\dg}_1}=2\lambda^2|l\rangle_{Y^{\dg}_1}-\lambda^2|1\rangle_{A^{\dg}_2}+\lambda^2|1\rangle_{B^{\dg}_1},\\
&\ml{H}|r\rangle_{Y^2}=2\lambda^2|r\rangle_{Y^2}-\lambda^2|L\rangle_{B^{\dg}_1}+\lambda^2|L\rangle_{A^{\dg}_2}.
\eea
We demand the proposed spin wave function to be an energy eigenstate:
\bea
\ml{H}|\psi_2(k)\rangle=E(k)|\psi_2(k)\rangle,
\eea
which leads to the following relations:
\begin{itemize}
  \item The bulk part ($x\neq 1,L$),
  \bea
  &2\lambda^2n(x)-\lambda^2n(x+1)-\lambda^2n(x-1)=E n(x),\label{EVE3}\\
  &2\lambda^2h(x)-\lambda^2h(x+1)-\lambda^2h(x-1)=E h(x),\label{EVE4}
  \eea
  which give the same dispersion relation
  \bea
  E(k)=2\lambda^2-2\lambda^2\cos k.
  \eea
  \item The left boundary part,
  \bea
  2\lambda^2 n(1)-\lambda^2 n(2)+\lambda^2\beta&=&En(1),\\
  2\lambda^2 h(1)-\lambda^2 h(2)-\lambda^2\beta&=&Eh(1),\\
  \lambda^2n(1)-\lambda^2h(1)+2\lambda^2\beta&=&E\beta.
  \eea
  Plugging eqs.(\ref{EVE3}) and (\ref{EVE4}) into these equations, we readily have
  \bea
  \beta&=&h(0),\\
  n(1)&=&-h(-1),\\
  n(0)&=&-h(0),
  \eea
  which means
  \bea
  &&N+\tilde{N}+H+\tilde{H}=0,\\
  &&Ne^{ik}+\tilde{N}e^{-ik}+He^{-ik}+\tilde{H}e^{ik}=0.
  \eea
  The equations are easily solved as
  \bea
  H=-\tilde{N},\quad N=-\tilde{H}.\label{eq3}
  \eea
  This gives the left reflection matrix in this section
  \bea
  K_l(k)
  =\left(\begin{array}{cc}0&-1\\
  -1&0\end{array}\right).
  \eea
  \item The right boundary part,
  \bea
  2\lambda^2n(L)-\lambda^2n(L-1)-\lambda^2\gamma&=&En(L),\\
  2\lambda^2h(L)-\lambda^2h(L-1)+\lambda^2\gamma&=&Eh(L),\\
  -\lambda^2n(L)+\lambda^2h(L)+2\lambda^2\gamma&=&E\gamma,
  \eea
  which imply
  \bea
  \gamma&=&n(L+1), \label{eq41}\\
  n(L+1)&=&-h(L+1),\\
  h(L)&=&-n(L+2).\label{eq43}
  \eea

  From these equations, we can get
  \bea H=-e^{-2ik(L+1)}\tilde{N}, \quad N=-e^{-2ik(L+1)}\tilde{H}.\eea
  Then the right reflection matrix in this sector is
  \bea
  K_r(k)
  =\left(\begin{array}{cc}0&-e^{-2ik}\\
  -e^{-2ik}&0\end{array}\right),
\eea
recalling the definition of right reflection matrix in eq.~(\ref{right}).

The compatibility of the eqs.~(\ref{eq41}-\ref{eq43}) with the solutions (\ref{eq3}) requires
  \bea
  e^{ik(2L+2)}=1,\quad k=\frac{n\pi}{L+1}, \quad n \in \mathbb{Z}.
  \eea
Therefore we get the Bethe equation of this sector and we also note that the effective length of the spin chain is $2L+2$ since two more boundary excitations participate in the interaction with the bulk excitations.

\end{itemize}
  \par The full left reflection matrix $K_l$ is then found to be
  \bea
  K_l(k)=
  \left
  (\begin{array}{cccc}0&0&0&-e^{-ik}\\0&0&-1&0\\
  0&-1&0&0\\
  -e^{-ik}&0&0&0
  \end{array}
  \right),
  \eea
  with the order of the excitations  as $A_2, B_1^\dg, A_2^\dg, B_1$. And the full right reflection matrix is
   \bea
  K_r(k)=
  \left
  (\begin{array}{cccc}0&0&0&-e^{-ik}\\0&0&-e^{-2ik}&0\\
  0&-e^{-2ik}&0&0\\
  -e^{-ik}&0&0&0
  \end{array}
  \right).
  \eea

  \par For the spin chain associated with the operator $\hat{O}'$, following the similar procedure shown above, we find the same reflection matrix after modifying the phase factor $e^{2ikL}$ in the definition of right reflection  matrix (\ref{right}) into $e^{ik(2L+1)}$. This modification is due to the  different effective length of the open spin chain. For the same reason, the Bethe equations for the two-particle and four-particle sectors will also be slightly modified. In order to prove the integrability of the Hamiltonian, we also need to know the bulk two-loop S-matrix which has been derived in \cite{Ahn:2009zg} using coordinate Bethe ansatz. We review this bulk S-matrix in appendix~\ref{Smatrix}. Equipped with the boundary and bulk scattering matrices, with the help of Mathematica program, we can verify the following reflection equations

\bea
&\left[S(k_1,k_2)\right]^{m_1m_2}_{l_1l_2}\left[K_l(k_2)\right]^{l_2}_{j_2}\left[S(-k_2,k_1)\right]^{l_1j_2}_{j_1i_2}\left[K_l(k_1)\right]^{j_1}_{i_1}\\\no
&=\left[K_l(k_1)\right]^{m_1}_{l_1}\left[S(-k_1,k_2)\right]^{l_1m_2}_{j_1l_2}\left[K_{l}(k_2)\right]^{l_2}_{j_2}\left[S(-k_2,-k_1)\right]^{j_1j_2}_{i_1i_2},
\eea
\bea
&\left[S(-k_1,-k_2)\right]^{m_1m_2}_{l_1l_2}\left[K_r(-k_1)\right]^{l_1}_{j_1}\left[S(-k_2,k_1)\right]^{j_1l_2}_{i_1j_2}\left[K_{r}(-k_2)\right]^{j_2}_{i_2}\\\no
&=\left[K_r(-k_2)\right]^{m_2}_{l_2}\left[S(-k_1,k_2)\right]^{m_1l_2}_{l_1j_2}\left[K_r(-k_1)\right]^{l_1}_{j_1}\left[S(k_2,k_1)\right]^{j_1j_2}_{i_1i_2},
\eea
  which, together with the validity of (bulk) YBE, guarantee the integrability of our open spin chain.

\section{Conclusion and discussions}
In this paper, we studied the two-loop integrability of planar ${\ml N}=3$ flavored ABJM theory in the scalar sector.
Rewriting the complete action in  a manifestly $SU(2)_R$ invariant way is the first step of the two-loop computation.
Working in 't Hooft limit, we only need to compute the ADMs of composite mesonic operators which naturally correspond to states on an  open alternating spin chain.
Taking the 't Hooft limit also tremendously simplifies the computation of the ADMs of this class of operators since the computation for the bulk part remians the same as
the one in ABJM theory. The final result of this computation can be re-expressed as a Hamiltonian on this open spin chain. It seemed hard to prove the integrability of this Hamiltonian by the algebraic Bethe ansatz method and for this reason we seek help from coordinate Bethe ansatz . We considered one-excitation states
and computed the left and right reflection matrices. Using these and the bulk two-loop S-matrix computed in \cite{Ahn:2009zg}, we verified the reflection equations for
both sides of the open spin chain. This established the two loop integrability of planar flavored ABJM theory in the scalar sector.

The immediate next step is to find the eigenvalues of the Hamiltonian.  For this, we need to solve an eigenvalue equation constructed from the S-matrices and the reflection matrices \cite{Schulz,Andrei:1982cr,Li:2013vdh}. To solve this eigenvalue equation,  off-diagonal
Bethe ansatz  (ODBA) \cite{WYCS} seems necessary here since the reflection matrices at both sides are non-diagonal \footnote{Notice that ODBA has been successfully applied to the open spin chain dual to open strings between two un-parallel branes in $AdS_5\times S^5$ \cite{Zhang:2015fea}.}. One may also try the algebraic Bethe ansatz from the beginning by solving the boundary Yang-Baxter equation obtained in this approach and analyze what kind of solution could reproduce the boundary Hamiltonian. We remind that nested coordinate Bethe ansatz  \cite{Schulz, Correa:2009dm} may be another choice as well.

One can also study the integrability of flavored ABJM theory in the Veneziano limit with $N_c, k, N_{f_1}, N_{f_2}\to \infty$ and $N_c/k, N_{f_1}/N_c, N_{f_2}/N_c$ fixed. In this case, the computation of the ADM will be much more complicated. For both single trace operators and mesonic operators,
some Feynman diagrams previously omitted due to $N_{f_i}/N_c$ suppression should be included now. And although the  mixing between certain single trace operators  and flavor-singlet mesonic operators like $\sum_{s=1}^{N_{f_2}}Y^\dagger_s XX^\dagger \cdots XX^\dagger Y^s$ and $\sum_{t=1}^{N_{f_1}}M^t X^\dagger X\cdots X^\dagger X M^\dagger_t$  is  $N_{f_i}/N_c$ suppressed in the 't Hooft limit, it should be taken into account in the Veneziano limit \cite{Gadde:2010zi}. Generally speaking, we need to consider the mixing among the generalized single trace operators involving $X, X^\dagger, \sum_{s=1}^{N_{f_2}}Y^s Y^\dagger_s, \sum_{t=1}^{N_{f_1}} M^\dagger_t M^t$ with the color indices in the final two composite operators un-contracted.

Another interesting question is that whether the planar integrability can be generalized to  the full sector of the theory and/or to higher orders of 't Hooft coupling constant $N_c/k$. One may even hope to pin down the all-order reflection matrix directly from the symmetry preserved by the vacuum as shown in  \cite{Hofman:2007xp} for a four-dimensional case. We leave all these important questions to future work.

\section*{Acknowledgments}
We would like to thank Junpeng Cao, Bin Chen, Yu Jia, Yunfeng Jiang, Yupeng Wang, Fakai Wen, Jia-ju Zhang for very helpful discussions. This work was in part supported by Natural Science Foundation of China under Grant Nos.  11425522(WY), 11434013(WY), 11575202(NB, HC, JW), 11475183(NB), 11275207(HC), 11690022(HC), 11305235(SH). The work of SH is also supported by Max-Planck fellowship in Germany. The research of MZ is partly supported by INFN Iniziativa Specifica ST$\&$FI.  NB, HC and JW thank Institute of Modern Physics, Northwest University  for warm hospitality in visits during this project. JW would also like to thank the participants of the advanced workshop ``Dark Energy and Fundamental Theory'' supported by the Special Fund for Theoretical Physics from NSFC with Grant No.~11447613 for stimulating discussions.

\begin{appendix}
\section{Some details of $\ml{N}=2$ formulation}\label{appendixa}
\subsection{The on-shell values of auxiliary fields}
The equations of motion for the auxiliary fields in chiral multiplets give
\bea
F^{\dg}_A&=&\frac{4\pi}{k}\left(W_AZW-WZW_A+W_ACE-BAW_A\right),\\
F^A&=&\frac{4\pi}{k}(-W^{\dg A}Z^{\dg}W^{\dg }+W^{\dg }Z^{\dg}W^{\dg A}-W^{\dg A}A^{\dg}B^{\dg }+E^{\dg }C^{\dg}W^{\dg A}),\\
G^{\dg A}&=&\frac{4\pi}{k}\left(ZWZ^A-Z^AWZ+CEZ^A-Z^ABA\right),\\
G_A&=&\frac{4\pi}{k}(-Z^{\dg}W^{\dg }Z^{\dg}_A+Z^{\dg}_AW^{\dg }Z^{\dg}-A^{\dg}B^{\dg }Z^{\dg}_A+Z^{\dg}_AE^{\dg }C^{\dg}),\\
J^{\dg}_s&=&\frac{4\pi}{k}(E_sCE+E_sZW),\\
J^s&=&\frac{4\pi}{k}(E^{\dg }C^{\dg}E^{\dg s}+W^{\dg }Z^{\dg}E^{\dg s}),\\
H^{\dg t}&=&\frac{4\pi}{k}(-A^{t}BA-A^{t}WZ),\\
H_t&=&\frac{4\pi}{k}(-A^{\dg}B^{\dg }A^{\dg}_t-Z^{\dg}W^{\dg }A^{\dg}_t),\\
K^{\dg s}&=&\frac{4\pi}{k}(CEC^s+ZWC^s),\\
K_s&=&\frac{4\pi}{k}(C^{\dg}_sE^{\dg }C^{\dg}+C^{\dg}_sW^{\dg }Z^{\dg}),\\
I^{\dg}_t&=&\frac{4\pi}{k}(-BAB_t-WZB_t),\\
I^t&=&\frac{4\pi}{k}(-B^{\dg t}A^{\dg}B^{\dg }-B^{\dg t}Z^{\dg}W^{\dg }).
\eea
The equations of motion for the auxiliary fields in gauge multiplets give
\bea
\sigma^n&=&\frac{2\pi}{k}\,\mbox{tr} T^n(ZZ^{\dg}-W^{\dg}W+CC^{\dg}-E^{\dg}E),\\
\hat{\sigma}^n&=&\frac{2\pi}{k}\,\mbox{tr}T^n(Z^{\dg}Z-WW^{\dg}+A^{\dg}A-BB^{\dg}),\\
\chi^n&=&-\frac{4\pi}{k}\,\mbox{tr}T^n(Z\zeta^{\dg}-\omega^{\dg}W+C\tau^{\dg}- v^{\dg}E),\\
\bar{\chi}^n&=&-\frac{4\pi}{k}\,\mbox{tr}T^n(\zeta Z^{\dg}-W^{\dg}\omega+\tau C^{\dg}-E^{\dg}v),\\
\hat{\chi}^n&=&-\frac{4\pi}{k}\,\mbox{tr}T^n(\zeta^{\dg}Z-W\omega^{\dg}-B\eta^{\dg}+\kappa^{\dg}A),\\
\hat{\bar{\chi}}^n&=&-\frac{4\pi}{k}\,\mbox{tr}T^n(Z^{\dg}\zeta-\omega W^{\dg}-\eta B^{\dg}+A^{\dg}\kappa),
\eea
where $T^n, n=1, \cdots, N_c^2$, is the generator of $U(N_c)$.
\subsection{The potential terms in $\ml{N}=2$ formulation}
The potentials from F-term and D-term contributions are given below
\bea
&&-V^{bos}_D\no\\\no
&=&-\frac{4\pi^2}{k^2}\,\mbox{tr}[(ZZ^{\dg}+W^{\dg}W+CC^{\dg}+E^{\dg}E)(ZZ^{\dg}-W^{\dg}W+CC^{\dg}-E^{\dg}E)\\\no
&&\times(ZZ^{\dg}-W^{\dg}W+CC^{\dg}-E^{\dg}E)]\\\no
&&-\frac{4\pi^2}{k^2}\,\mbox{tr}[(Z^{\dg}Z+WW^{\dg}+A^{\dg}A+BB^{\dg})(Z^{\dg}Z-WW^{\dg}+A^{\dg}A-BB^{\dg})\\\no
&&\times (Z^{\dg}Z-WW^{\dg}+A^{\dg}A-BB^{\dg})]\\\no
&&+\frac{8\pi^2}{k^2}\,\mbox{tr}\left[Z^{\dg}_A(ZZ^{\dg}-W^{\dg}W+CC^{\dg}-E^{\dg}E)Z^A(Z^{\dg}Z-WW^{\dg}+A^{\dg}A-BB^{\dg})\right]\\
&&+\frac{8\pi^2}{k^2}\,\mbox{tr}\left[W^{\dg A}(Z^{\dg}Z-WW^{\dg}+A^{\dg}A-BB^{\dg})W_A(ZZ^{\dg}-W^{\dg}W+CC^{\dg}-E^{\dg}E)\right],
\eea
\bea
&&\no-V_F^{bos}\\&=&-\frac{16\pi^2}{k^2}\,\mbox{tr}\,\left(-W_AZW+WZW_A-W_ACE+BAW_A\right)(W^{\dg A}Z^{\dg}W^{\dg }-W^{\dg }Z^{\dg}W^{\dg A}\no\\\no
&&+W^{\dg A}A^{\dg}B^{\dg }-E^{\dg }C^{\dg}W^{\dg A})-\frac{16\pi^2}{k^2}\,\mbox{tr}\,\left(ZWZ^A-Z^AWZ+CEZ^A-Z^ABA\right)\\
&&(-Z^{\dg}W^{\dg }Z^{\dg}_A+Z^{\dg}_AW^{\dg }Z^{\dg}-A^{\dg}B^{\dg }Z^{\dg}_A+Z^{\dg}_AE^{\dg }C^{\dg})-\frac{16\pi^2}{k^2}\,\mbox{tr}\,(E_sCE+E_sZW)\\\no
&&(E^{\dg }C^{\dg}E^{\dg s}+W^{\dg }Z^{\dg}E^{\dg s})-\frac{16\pi^2}{k^2}\,\mbox{tr}\,(A^{t}BA+A^{t}WZ)(A^{\dg}B^{\dg }A^{\dg}_t+Z^{\dg}W^{\dg }A^{\dg}_t)\\\no
&&-\frac{16\pi^2}{k^2}\,\mbox{tr}\,(CEC^s+ZWC^s)(C^{\dg}_sE^{\dg }C^{\dg}+C^{\dg}_sW^{\dg }Z^{\dg})-\frac{16\pi^2}{k^2}\,\mbox{tr}\,(BAB_t+WZB_t)\\\no
&&(B^{\dg t}A^{\dg}B^{\dg }+B^{\dg t}Z^{\dg}W^{\dg }),
\eea
\bea
-V^{ferm}_D\no&=&-\frac{2\pi i}{k}\,\mbox{tr}(\zeta\zeta^{\dg}+\tau\tau^{\dg}-\omega^{\dg}\omega-v^{\dg}v)(ZZ^{\dg}-W^{\dg}W+CC^{\dg}-E^{\dg}E)\\
&&+\frac{2\pi i}{k}\,\mbox{tr}(\zeta^{\dg}\zeta-\omega\omega^{\dg}-\eta\eta^{\dg}+\kappa^{\dg}\kappa)(Z^{\dg}Z-WW^{\dg}+A^{\dg}A-BB^{\dg})\\\no
&&-\frac{4\pi i}{k}\,\mbox{tr}(\zeta Z^{\dg}-W^{\dg}\omega+\tau C^{\dg}-E^{\dg} v)(Z\zeta^{\dg}-\omega^{\dg}W+C\tau^{\dg}-v^{\dg}E)\\\no
&&+\frac{4\pi i}{k}\,\mbox{tr}(Z^{\dg}\zeta-\omega W^{\dg}+A^{\dg}\kappa-\eta B^{\dg})(\zeta^{\dg}Z-W\omega^{\dg}+\kappa^{\dg}A-B\eta^{\dg}),
\eea
\bea
-V^{ferm}_F&=&\mbox{tr}\left(\frac{2\pi}{k}\ep_{AC}\ep^{BD}[-2\zeta^A W_B Z^C \omega_D-2\zeta^A \omega_B Z^C W_D-Z^A\omega_BZ^C\omega_D-\zeta^AW_B\zeta^CW_D]\right.\\\no
&&-\frac{2\pi}{k}\ep^{AC}\ep_{BD}[2\zeta^{\dg}_AW^{\dg B}Z^{\dg}_C\omega^{\dg D}+2\zeta^{\dg}_A\omega^{\dg B}Z^{\dg}_CW^{\dg D}+Z^{\dg}_A\omega^{\dg B}Z^{\dg}_C\omega^{\dg D}+\zeta^{\dg}_AW^{\dg B}\zeta^{\dg}_CW^{\dg D}]\\\no
&&-\frac{2\pi}{k}[-2\tau v C E-2\tau E C v-\tau E \tau E-C v C v]\\\no
&&+\frac{2\pi}{k}[-2\eta \kappa B A-2B \kappa \eta A-B\kappa B \kappa-\eta A\eta A]\\\no
&&+\frac{2\pi}{k}[2\kappa^{\dg}\eta^{\dg}A^{\dg}B^{\dg}+2\kappa^{\dg}B^{\dg}A^{\dg}\eta^{\dg}+\kappa^{\dg}B^{\dg}\kappa^{\dg}B^{\dg}+A^{\dg}\eta^{\dg}A^{\dg}\eta^{\dg}]\\\no
&&-\frac{2\pi}{k}[2v^{\dg}\tau^{\dg}E^{\dg}C^{\dg}+2E^{\dg}\tau^{\dg}v^{\dg}C^{\dg}+E^{\dg}\tau^{\dg}E^{\dg}\tau^{\dg}+v^{\dg}C^{\dg}v^{\dg}C^{\dg}]\\\no
&&+\frac{4\pi}{k}[ZW\tau v+{{\zeta \omega C E}}+Z\omega C v+Z\omega \tau E+\zeta W C v+\zeta W\tau E]\\\no
&&+\frac{4\pi}{k}[-WZ\eta \kappa-{{\omega\zeta B A}}-W\zeta B\kappa-W\zeta\eta A-\omega ZB\kappa-\omega Z \eta A]\\\no
&&+\frac{4\pi}{k}[{{A^{\dg}B^{\dg}\zeta^{\dg}\omega^{\dg}}}+A^{\dg}\eta^{\dg}Z^{\dg}\omega^{\dg}+A^{\dg}\eta^{\dg}\zeta^{\dg}W^{\dg}+\kappa^{\dg}B^{\dg}Z^{\dg}\omega^{\dg}+\kappa^{\dg}B^{\dg}\zeta^{\dg}W^{\dg}+\kappa^{\dg}\eta^{\dg}Z^{\dg}W^{\dg}]\\\no
&&\left.+\frac{4\pi}{k}[{{-E^{\dg}C^{\dg}\omega^{\dg}\zeta^{\dg}}}-E^{\dg}\tau^{\dg}W^{\dg}\zeta^{\dg}-E^{\dg}\tau^{\dg}\omega^{\dg}Z^{\dg}
-v^{\dg}C^{\dg}W^{\dg}\zeta^{\dg}-v^{\dg}C^{\dg}\omega^{\dg}Z^{\dg}-v^{\dg}\tau^{\dg}W^{\dg}Z^{\dg}]\right),
\eea
where the summation indices are suppressed for those obvious contractions between two adjacent fields.

\section{BPS property of the reference state}\label{susy}
For the supersymmetry transformation of ${\cal N}=3$ Chern-Simon-matter theories, we follow the convention of \cite{Ouyang:2015bmy}.\footnote{Here we only consider the Poincare supercharges and neglect the conformal supercharges.} We perform a Wick rotation to three dimensional Euclidean space.

The supersymmetry transformations of $Y^\dg_i, Y^i, X^\dg_{iA}, X^{iA}$ are
\bea \delta Y^\dg_i&=&\ii \psi_{j}^\dg\theta^j_{\,i}, \\
\delta Y^i&=&\ii \theta^i_{\,j}\psi^j,\\
\delta X^\dg_{iA}&=&\ii \xi_{jA}^\dg \theta^j_{\,i},\\
\delta X^{iA}&=&\ii \theta^i_{\,j} \xi^{jA},
\eea
where the supersymmetry transformation parameters $\theta^i_{\,j}$ satisfy the constraint
$\theta^i_{\,i}=0$. It is easy to see that the vacuum state
\bea
|\Omega\rangle=|Y^{\dg}_2X^{11}X^{\dg}_{22}\cdots X^{11}X^{\dg}_{22}Y^1\rangle
\eea
is invariant under the supersymmetry transformation with $\theta^1_{\,1}=\theta^2_{\,2}=\theta^1_{\,2}=0$, so it is
$1/3$-BPS.

\section{The bulk S-matrix}\label{Smatrix}
In this appendix we briefly review the bulk S-matrix computed in \cite{Ahn:2009zg}. Our convention is that in  $S_{I_1I_2}^{J_1J_2}$,
$I_i$ is used to denote the in-state of the $i$-th particle and $J_i$ is for the out-state of the $i$-th particle.

We define \be u_i\equiv \frac12\cot\frac{k_i}2, \quad u_{ij}\equiv u_i-u_j. \ee
The non-zero elements of the bulk S-matrix is
\bea S^{\phi\phi}_{\phi\phi}(k_2, k_1)=\frac{u_{21}+i}{u_{21}-i}, \eea
where $\phi$ is one of $A_2, B_1^\dagger, A_2^\dagger, B_1$;
\bea S^{A_2B_1^\dg}_{A_2B_1^\dg}(k_2, k_1)=S^{B_1^\dg A_2}_{B_1^\dg A_2}(k_2, k_1)=S^{A_2^\dg B_1}_{A_2^\dg B_1}(k_2, k_1)=
S^{B_1A_2^\dg}_{B_1A_2^\dg }(k_2, k_1)=\frac{u_{21}}{u_{21}-i}; \eea
\bea S^{B_1^\dg A_2}_{A_2B_1^\dg}(k_2, k_1)=S^{A_2B_1^\dg }_{B_1^\dg A_2}(k_2, k_1)=S^{B_1A_2^\dg }_{A_2^\dg B_1}(k_2, k_1)=
S^{A_2^\dg B_1}_{B_1A_2^\dg }(k_2, k_1)=\frac{i}{u_{21}-i}; \eea
\bea  S^{A_2B_1}_{A_2B_1}(k_2, k_1)=S^{B_1 A_2}_{B_1 A_2}(k_2, k_1)=S^{A_2^\dg B_1^\dg}_{A_2^\dg B_1^\dg}(k_2, k_1)=
S^{B_1^\dg A_2^\dg}_{B_1^\dg A_2^\dg }(k_2, k_1)=1; \eea

\bea S^{A_2A_2^\dg}_{A_2A_2^\dg}(k_2, k_1)=S^{B_1^\dg B_1}_{B_1^\dg B_1}(k_2, k_1)=S^{A_2^\dg A_2}_{A_2^\dg A_2}(k_2, k_1)=
S^{B_1B_1^\dg}_{B_1B_1^\dg }(k_2, k_1)=\frac{u_{12}}{u_{12}-i}; \eea

\bea S^{A_2A_2^\dg}_{B_1^\dg B_1}(k_2, k_1)=S^{B_1^\dg B_1}_{A_2A_2^\dg}(k_2, k_1)=S^{A_2^\dg A_2}_{B_1B_1^\dg}(k_2, k_1)=
S^{B_1B_1^\dg}_{A_2^\dg A_2}(k_2, k_1)=\frac{i}{u_{12}-i}. \eea

We also verified that this S-matrix satisfies the YBE.

 \end{appendix}


\begin{thebibliography}{}

\bibitem{Brink:1976bc}
  L.~Brink, J.~H.~Schwarz and J.~Scherk,
  ``Supersymmetric Yang-Mills Theories,''
  Nucl.\ Phys.\ B {\bf 121}, 77 (1977),  doi:10.1016/0550-3213(77)90328-5.

\bibitem{ABJM}
  O.~Aharony, O.~Bergman, D.~L.~Jafferis and J.~Maldacena,
  ``$\ml{N}=6$ superconformal Chern-Simons-matter theories, M2-branes and their gravity duals,''
  JHEP {\bf 0810}, 091 (2008)
  doi:10.1088/1126-6708/2008/10/091
  [arXiv:0806.1218 [hep-th]].


  \bibitem{Beisert:2010jr}
  N.~Beisert, C.~Ahn, L.~F.~Alday, Z.~Bajnok, J.~M.~Drummond, L.~Freyhult, N.~Gromov and R.~A.~Janik {\it et al.},
  ``Review of AdS/CFT Integrability: An Overview,''
  Lett.\ Math.\ Phys.\  {\bf 99}, 3 (2012),
  doi:10.1007/s11005-011-0529-2,
  [arXiv:1012.3982 [hep-th]].


\bibitem{Minahan:2002}
  J.~A.~Minahan and K.~Zarembo,
  ``The Bethe ansatz for ${\ml N}=4$ super Yang-Mills,''
  JHEP {\bf 0303}, 013 (2003),
 doi:10.1088/1126-6708/2003/03/013,
  [hep-th/0212208].


\bibitem{Minahan:2008}
J.~A.~Minahan and K.~Zarembo,
``The Bethe ansatz for superconformal Chern-Simons,''
JHEP {\bf{0809}}, 400 (2008),  doi:10.1088/1126-6708/2008/09/040, [arXiv:0806.3951 [hep-th]].

\bibitem{Bak:2008}
D.~Bak and S.~J.~Rey,
``Integrable Spin Chain in Superconformal Chern-Simons Theory,''
JHEP {\bf{0810}}, 053 (2008),   doi:10.1088/1126-6708/2008/10/053, [arXiv:0807.2063 [hep-th]].


\bibitem{Erler:2005nr}
  T.~Erler and N.~Mann,
  ``Integrable open spin chains and the doubling trick in ${\ml N}=2$ SYM with fundamental matter,''
  JHEP {\bf 0601}, 131 (2006),
  doi:10.1088/1126-6708/2006/01/131,
  [hep-th/0508064].

\bibitem{Chen:2004mu}
  B.~Chen, X.~J.~Wang and Y.~S.~Wu,
  ``Integrable open spin chain in super Yang-Mills and the plane wave / SYM duality,''
  JHEP {\bf 0402}, 029 (2004),
  doi:10.1088/1126-6708/2004/02/029,
  [hep-th/0401016].

 \bibitem{Chen:2004yf}
  B.~Chen, X.~J.~Wang and Y.~S.~Wu,
  ``Open spin chain and open spinning string,''
  Phys.\ Lett.\ B {\bf 591}, 170 (2004),
  doi:10.1016/j.physletb.2004.04.013,
  [hep-th/0403004].


\bibitem{Hohenegger:2009}
S.~Hohenegger and I.~Kirsch,
``A note on the holography of
Chern-Simons matter theories with flavour,''
JHEP {\bf{0904}}, 129 (2009), doi:10.1088/1126-6708/2009/04/129, [arXiv:0903.1730 [hep-th]].

\bibitem{Gaiotto:2009}
D.~Gaiotto and D.~L.~Jafferis,
``Notes on adding D6 branes wrapping $\mathbb{RP}^3$ in $AdS_4\times \mathbb{CP}^3$,''
JHEP {\bf{11}} (2012) 015, doi:10.1007/JHEP11(2012)015, [arXiv:0903.2175 [hep-th]].

\bibitem{Hikida:2009}
Y.~Hikida, W.~Li, and T.~Takayanagi,
``ABJM with Flavors and FQHE,''
JHEP {\bf{07}} (2009) 065,  doi:10.1088/1126-6708/2009/07/065, [arXiv:0903.2194 [hep-th]].

\bibitem{Bak:2009mq}
  D.~Bak, H.~Min and S.~J.~Rey,
  ``Generalized Dynamical Spin Chain and $4$-Loop Integrability in ${\ml N}=6$ Superconformal Chern-Simons Theory,''
  Nucl.\ Phys.\ B {\bf 827}, 381 (2010),
  doi:10.1016/j.nuclphysb.2009.10.011,
  [arXiv:0904.4677 [hep-th]].

\bibitem{Berenstein:2005vf}
  D.~Berenstein and S.~E.~Vazquez,
  ``Integrable open spin chains from giant gravitons,''
  JHEP {\bf 0506}, 059 (2005)
  doi:10.1088/1126-6708/2005/06/059
  [hep-th/0501078].

\bibitem{Ahn:2009zg}
  C.~Ahn and R.~I.~Nepomechie,
  ``Two-loop test of the $\ml{N}=6$ Chern-Simons theory S-matrix,''
  JHEP {\bf 0903}, 144 (2009),
  doi:10.1088/1126-6708/2009/03/144,
  [arXiv:0901.3334 [hep-th]].

\bibitem{Ahn:2008aa}
  C.~Ahn and R.~I.~Nepomechie,
  ``${\ml N}=6$ super Chern-Simons theory S-matrix and all-loop Bethe ansatz equations,''
  JHEP {\bf 0809}, 010 (2008),
  doi:10.1088/1126-6708/2008/09/010,
  [arXiv:0807.1924 [hep-th]].

\bibitem{DeWolfe:2004}
O.~DeWolfe and N.~Mann,
``Integrable Open Spin Chains in Defect Conformal Field Theory,''
JHEP {\bf{0404}}, 035 (2004), doi:10.1088/1126-6708/2004/04/035,  [hep-th/0401041].

\bibitem{Benna:2008}
M.~Benna, I.~Klebanov, T.~Klose and M.~Smedback,
``Superconformal Chern-Simons theories and $AdS_4/CFT_3$ correspondence,''
JHEP {\bf{0809}} (2008) 072,  doi:10.1088/1126-6708/2008/09/072,  [arXiv:0806.1519 [hep-th]].

\bibitem{Benna:2009}
M.~Benna, I.~Klebanov, and T.~Klose,
``Charges of Monopole Operators in Chern-Simons Yang-Mills Theory,''
JHEP {\bf{01}} (2010) 110, doi:10.1007/JHEP01(2010)110, [arXiv:0906.3008 [hep-th]].

\bibitem{He:2013hxd}
  S.~He and J.~B.~Wu,
  ``Note on Integrability of Marginally Deformed ABJ(M) Theories,''
  JHEP {\bf 1304}, 012 (2013),
  Erratum: [JHEP {\bf 1604}, 139 (2016)],
  doi:10.1007/JHEP04(2013)012, 10.1007/JHEP04(2016)139,
  [arXiv:1302.2208 [hep-th]].

\bibitem{Chen:2016geo}
  H.~H.~Chen, P.~Liu and J.~B.~Wu,
  ``Y-system for $\gamma$-deformed ABJM Theory,''
  JHEP03(2017)133, doi:10.1007/JHEP03(2017)133
  [arXiv:1611.02804 [hep-th]].
\bibitem{Li:2013vdh}
  Y.~Y.~Li, J.~Cao, W.~L.~Yang, K.~Shi and Y.~Wang,
  ``Exact solution of the one-dimensional Hubbard model with arbitrary boundary magnetic fields,''
  Nucl.\ Phys.\ B {\bf 879}, 98 (2014),
  doi:10.1016/j.nuclphysb.2013.12.004,
  [arXiv:1311.0432 [cond-mat.str-el]].

  \bibitem{WYCS}
  Y.~Wang, W.-L.~Yang,  J.~Cao and K.~Shi, ``Off-Diagonal Bethe Ansatz for Exactly Solvable Models,'' Springer Press, 2015.

\bibitem{Zhang:2015fea}
  X.~Zhang, J.~Cao, S.~Cui, R.~I.~Nepomechie, W.~L.~Yang, K.~Shi and Y.~Wang,
  ``Bethe ansatz for an AdS/CFT open spin chain with non-diagonal boundaries,''
  JHEP {\bf 1510}, 133 (2015),
  doi:10.1007/JHEP10(2015)133,
  [arXiv:1507.08866 [hep-th]].

\bibitem{Schulz}
H.~Schulz, ``Hubbard chain with reflecting ends,'' J.\ Phys.\ C \ {\bf 18} (1985) 581, doi:10.1088/0022-3719/18/3/010.

\bibitem{Andrei:1982cr}
  N.~Andrei, K.~Furuya and J.~H.~Lowenstein,
  ``Solution of the Kondo Problem,''
  Rev.\ Mod.\ Phys.\  {\bf 55}, 331 (1983),
  doi:10.1103/RevModPhys.55.331.

\bibitem{Correa:2009dm}
  D.~H.~Correa and C.~A.~S.~Young,
  ``Asymptotic Bethe equations for open boundaries in planar AdS/CFT,''
  J.\ Phys.\ A {\bf 43}, 145401 (2010),
  doi:10.1088/1751-8113/43/14/145401,
  [arXiv:0912.0627 [hep-th]].

\bibitem{SW}
M.~Shiroishi and M.~Wadati, ``Integrable boundary conditions for one-dimensional Hubbard model'',
J.\ Phys.\ Soc.\ Jpn {\bf 66} (1997), 2288, 10.1143/JPSJ.66.2288, [arXiv:cond-mat/9708011].


\bibitem{Gadde:2010zi}
  A.~Gadde, E.~Pomoni and L.~Rastelli,
  ``Spin Chains in ${\ml N}=2$ Superconformal Theories: From the $Z_2$ Quiver to Superconformal QCD,''
  JHEP {\bf 1206}, 107 (2012),
  doi:10.1007/JHEP06(2012)107,
  [arXiv:1006.0015 [hep-th]].

\bibitem{Hofman:2007xp}
  D.~M.~Hofman and J.~M.~Maldacena,
  ``Reflecting magnons,''
  JHEP {\bf 0711}, 063 (2007),
  doi:10.1088/1126-6708/2007/11/063,
  [arXiv:0708.2272 [hep-th]].

\bibitem{Ouyang:2015bmy}
  H.~Ouyang, J.~B.~Wu and J.~j.~Zhang,
  ``Construction and classification of novel BPS Wilson loops in quiver Chern-Simons-matter theories,''
  Nucl.\ Phys.\ B {\bf 910}, 496 (2016),
  doi:10.1016/j.nuclphysb.2016.07.018,
  [arXiv:1511.02967 [hep-th]].




\end{thebibliography}
\end{document}